\documentclass[12pt]{article}
\usepackage[utf8]{inputenc}
\usepackage[english]{babel}
\usepackage{authblk}
\usepackage{graphicx}
\usepackage{mathptmx}
\usepackage[singlespacing]{setspace}
\usepackage[headheight=1in,margin=1in]{geometry}
\usepackage{fancyhdr}

\graphicspath{{images/}}

\title{The ``Non-Musk Effect'' at Twitter}
\author[]{Dmitry Zinoviev, Arkapravo Sarkar, Pelin Bicen\\\vspace{-0.5em}
  Suffolk University, Boston, MA, USA}
 
\date{March 2023}

\begin{document}

\maketitle

\begin{abstract}  
 Elon Musk has long been known to significantly impact Wall Street through his controversial statements and actions, particularly through his own use of social media. An innovator and visionary entrepreneur, Musk is often considered a poster boy for all entrepreneurs worldwide. It is, thus, interesting to examine the effect that Musk might have on Main Street, i.e., on the social media activity of other entrepreneurs. In this research, we study and quantify this ``Musk Effect,'' i.e., the impact of Musk's recent and highly publicized acquisition of Twitter on the tweeting activity of entrepreneurs. Using a dataset consisting of 9.94 million actual tweets from 47,190 self-declared entrepreneurs from seven English-speaking countries (US, Australia, New Zealand, UK, Canada, South Africa, and Ireland) spanning 71 weeks and encompassing the entire period from the rumor that Musk may buy Twitter till the completion of the acquisition, we find that only about 2.5\% of the entrepreneurs display a significant change in their tweeting behavior over the time.

We believe that our study is one of the first works to examine the effect of Musk's acquisition of Twitter on the actual tweeting behavior of Twitter users (entrepreneurs). By quantifying the impact of the Musk Effect on Main Street, we provide a comparison with the effect Musk's actions have on Wall Street. Finally, our systematic identification of the characteristics of entrepreneurs most affected by the Musk Effect has practical implications for academics and practitioners alike.

 \vskip\baselineskip%
\end{abstract}

\section{Introduction\label{sec:intro}}

Elon Musk, the CEO of Tesla and SpaceX, is known for his quirky personality and controversial statements. His actions have often caused significant movements in the stock market, especially in the tech industry. However, the impact of Musk's behavior and actions on social media, specifically on other entrepreneurs, has yet to be studied extensively. This research seeks to fill this gap by examining the ``Musk Effect'' on the Twitter activity of self-declared entrepreneurs.

The timeline surrounding Musk's recent and highly publicized acquisition of Twitter is worth noting. On November 29, 2021, Jack Dorsey stepped down as Twitter CEO. Shortly after, on January 31, 2022, Musk began purchasing Twitter stock. On April 14, 2022, Musk made an unsolicited and nonbinding offer to purchase the company, which caused significant speculation and buzz on social media platforms. However, on July 8, 2022, Musk announced his intention to terminate the proposed acquisition, causing a stir in the financial markets. Finally, on October 27, 2022, Musk and Twitter closed the deal, bringing his vision of a decentralized social media platform to life.

Against this backdrop, our study quantifies the ``Musk Effect'' on the tweeting activity of entrepreneurs from seven English-speaking countries (US, Australia, New Zealand, UK, Canada, South Africa, and Ireland) for 71 weeks. For this study, we downloaded and created a dataset of 9.9 million tweets from more than 47,000 self-declared entrepreneurs. We performed a detailed analysis of all the tweets, including the account details like username, account age, number of lists, posts, followers, following, and any hashtags and mentions associated with every tweet. Using trajectory cluster analysis, we could classify the entrepreneurs into eight broad non-overlapping groups (clusters). Using hashtag-based analysis, we extracted the popular conversation topics from these entrepreneurs' tweets. Further, with the entrepreneurs' following information, i.e., based on whom each of our entrepreneurs follows on Twitter, we could estimate and plot the political leaning of these entrepreneurs. Finally, we used multivariate logistic regression to determine some identifying characteristics that could be attributed to each of the entrepreneur clusters that emerged from our analysis.

The rest of the paper is structured as follows. The second section details the steps in our data acquisition methodology and our final sample selection criterion. The third section presents a comprehensive data analysis, including a detailed description of the user classification methodology and political leaning estimation. The fourth section discusses the regression analysis and the related findings for each cluster of entrepreneurs, and the final section concludes the paper.   

In the remainder of this paper, vertical dashed lines in the figures represent the following events:
\begin{itemize}
\item January 31, 2022: Elon Musk began purchasing Twitter stock (in some figures).
\item April 14, 2022: Elon Musk made an offer to Twitter to purchase the company.
\item July 8, 2022: Elon Musk announced his intention to terminate the acquisition.
\item October 27, 2022: Elon Musk and Twitter closed the deal.
\end{itemize}

\section{Data Acquisition}

Our dataset comprises 9,943,721 tweets posted by 47,189 unique users for 71 weeks, commencing October 1, 2022, and concluding on February 8, 2023. The users were randomly selected from the live Twitter stream based on two criteria. Firstly, they must originate from an officially English-speaking country (the USA, Great Britain, Canada, Ireland, New Zealand, or Australia) or South Africa. Secondly, their Twitter profile descriptions must contain at least one of the following keywords, either verbatim or as a hashtag: ``entrepreneur,'' ``owner,'' ``founder,'' ``co-founder,'' ``business owner,'' ``CEO,'' ``co-owner,'' ``representative,'' ``consultant,'' ``assistant,'' ``analyst,'' ``officer,'' ``partner,'' ``developer,'' ``head,'' ``coordinator,'' ``specialist,'' ``executive,'' ``manager,'' ``managing director,'' ``chief executive,'' ``director,'' and ``VP''~\cite{hanik2023}. We collected each user's identifier, name, and username, along with basic public statistics such as account age and the number of lists, posts, followers, and following. Furthermore, we recorded whether their accounts had URLs and were verified. For every tweet in the dataset, we have collected its unique identifier as well as the unique identifier of its author. Furthermore, we have recorded the date of posting and the original text, including all mentions (``@'') and hashtags (``\#'').

In our study, we utilized a series of simple heuristics based on gendered names, usernames, pronouns, family roles, and occupations to identify the users' genders, where feasible. Subsequently, we employed gendered accounts to train a logistic regression algorithm for predicting the gender of remaining users. Our model exhibited a precision of 99.3\%, thus rendering us confident in our acceptance of the predicted genders~\cite{hanik2023}.

Due to the limited availability of tweets (at most 3,000 most recent tweets per account, including retweets), we could only obtain partial timelines for some highly active users, with most posts made in the last weeks of the observation period. This disparity created a distorted perception of increasing posting frequency over time. Since we aimed to employ weekly posting frequency to gauge Twitter's response to Musk's acquisition, such bias was untenable.

\begin{figure}[tb!]
\centering
\includegraphics[width=0.8\textwidth]{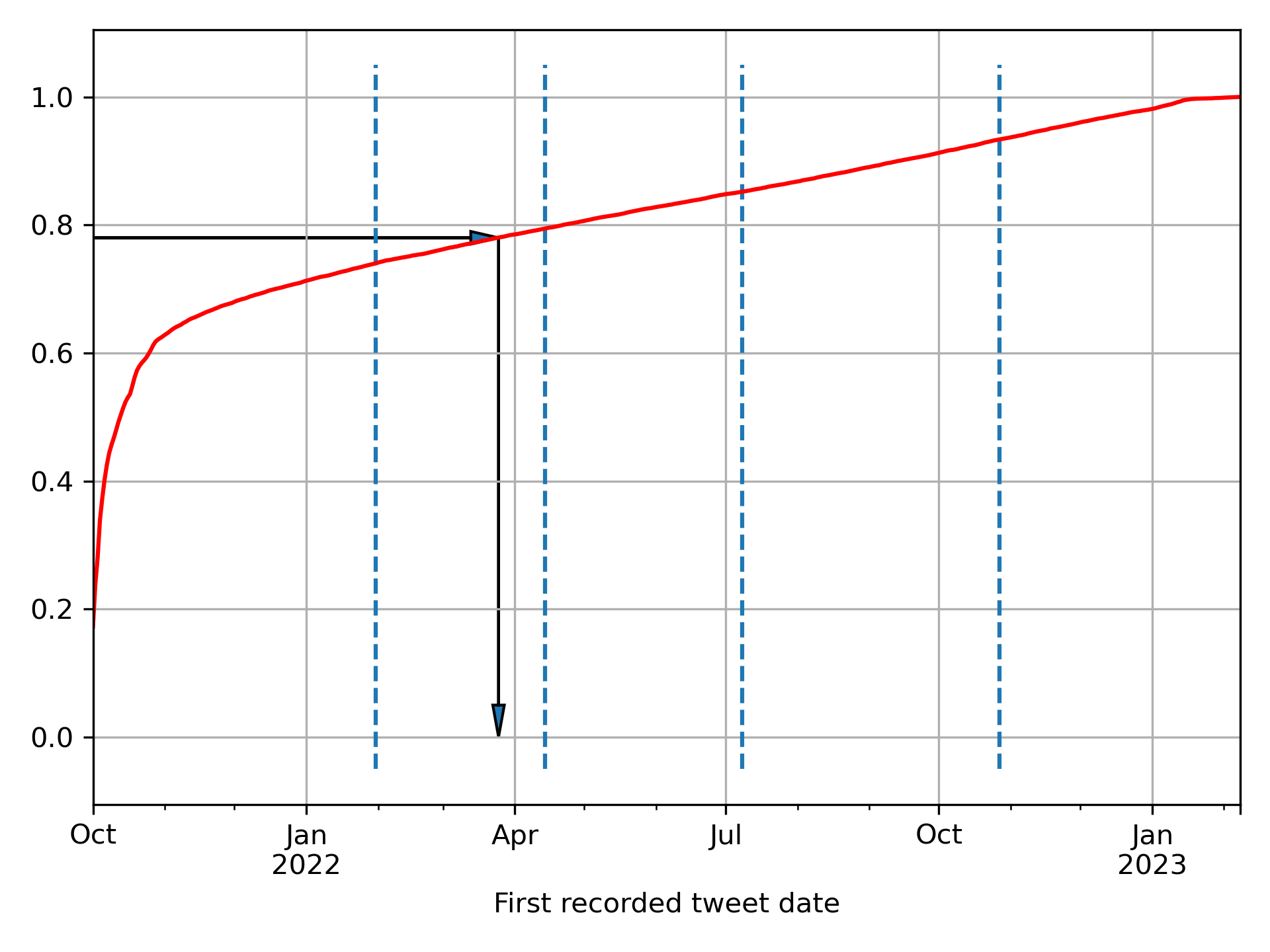}
\caption{Cumulative distribution of the dates of the first recorded tweet. See Section~\ref{sec:intro} for the explanation of the vertical dashed lines. Notably, we possess complete timelines for the accounts recorded on or after March 25, 2022, corresponding to 78\% of the dataset, as indicated by the black arrows.}
\label{fig:too_active}
\end{figure}

To examine and mitigate this bias, we plotted the cumulative distribution of the dates of the first recorded tweet (Figure~\ref{fig:too_active}). For instance, as illustrated by the black arrows in the figure, accounts with complete and uninterrupted timelines originating on or before March 25, 2022, well before any overt acquisition-related Musk activity, constituted 78\% of all harvested accounts, as shown by the black arrows in the figure. Consequently, if we sacrificed 22\% of the dataset, we could assure timeline uniformity. We could expand the timeline at the expense of the number of users or include more active users at the cost of a shorter timeline, as suggested by the chart. As previously mentioned, we ultimately opted for a subset of 78\% of users whose timelines began on or before March 25, 2022. The chosen observation window encompassed the entire Twitter acquisition saga, including its prelude.

\begin{table}[!b]\centering
  \caption{\label{table:users_by_country}Number of unique users per country in the truncated dataset}\vskip0.5\baselineskip
  \begin{tabular}{lr|lr|lr|lr}\hline
    Country & Count &Country & Count &Country & Count&Country & Count  \\\hline
    AU & 3,732 & CA &  5,453 & IE & 2,866 & NZ &   812 \\ 
    UK & 4,561 & US & 13,918 & ZA & 5,486 &    &       \\\hline
  \end{tabular}
\end{table}

The truncated dataset contains 4,034,007 (40.5\%) tweets posted by 36,828 (78\%) unique users throughout 46 (65\%) weeks. Of the users, 2,712 did not post anything during the measurement period. As a result, the number of users subjected to further analysis is 34,116 (72\%). To summarize the composition of the dataset, Table~\ref{table:users_by_country} provides the number of unique users per country of tweet origin.

\section{Data Analysis}

\subsection{Analyzing Tweet and Hashtag Frequencies}

\begin{figure}[tb!]
\centering
\includegraphics[width=\textwidth]{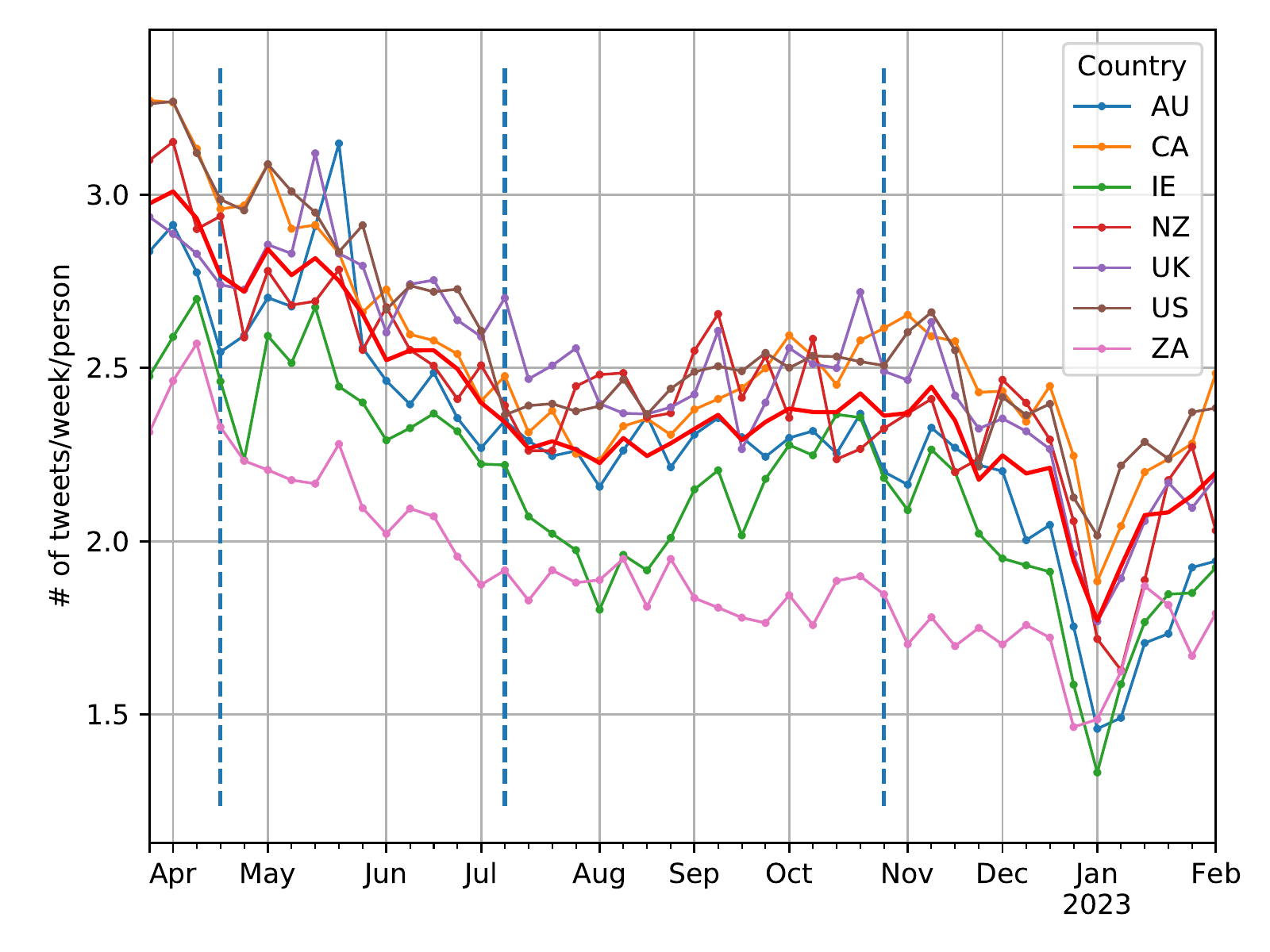}
\caption{Tweeting frequency, per country, over time. The red line shows the mean tweeting frequency.}
\label{fig:dynamics}
\end{figure}

Figure~\ref{fig:dynamics} presents the mean tweeting frequency, measured as the number of tweets per week per person for each country of origin, and the mean tweeting frequency for the entire dataset. The noticeable and consistent decline in tweeting frequency across all countries on the right side of the chart is attributed to the Christmas/New Year holiday season, with the decline being less pronounced in South Africa. 

Additionally, the frequencies display an almost continuous decline between April and early July, which will be explored further in Section~\ref{sec:classification} upon introducing Figure~\ref{fig:tag_trends}.

This study examines the user perception of Elon Musk's online persona on Twitter. Users have the option to refer to Mr. Musk through either the mention ``@elonmusk'' or the hashtag ``\#elonmusk.'' The latter option is considered indirect, as it does not immediately notify Mr.~Musk of the tweet's contents; it is primarily intended to contribute to the trending activity on the site's homepage and attract the attention of other subscribers~\cite{bruns2014structural,burgess2020}. 

\begin{figure}[tb!]
\centering
\includegraphics[width=\textwidth]{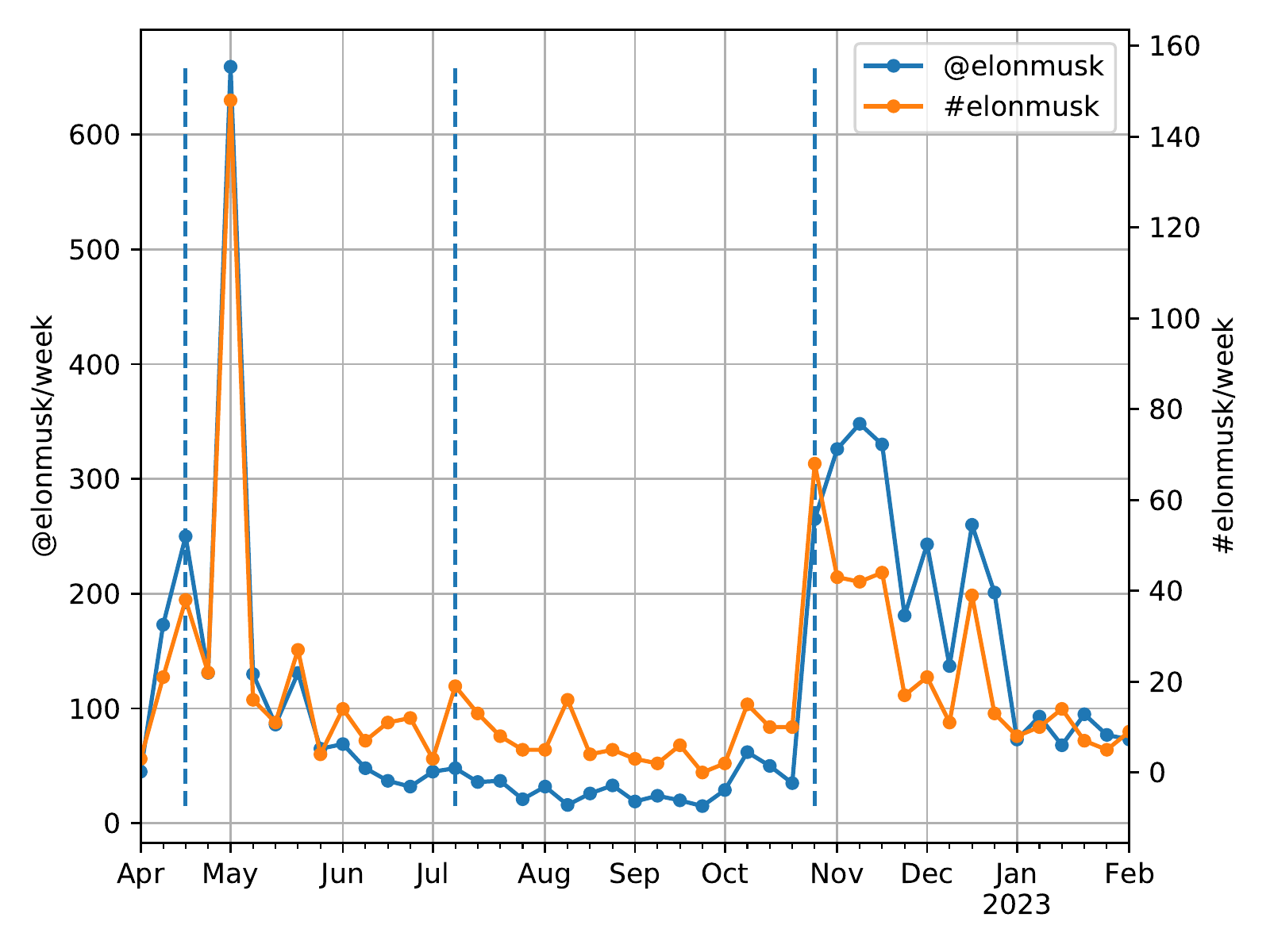}
\caption{Mentions of Elon Musk (@elonmusk) and the namesake hashtags (\#elonmusk) over time.}
\label{fig:musk_refs}
\end{figure}

On the other hand, a mention option is a direct form of communication that immediately alerts Mr.~Musk or his designated account handler of the mention. As such, it may be viewed as a more personal approach to garner the direct recipient's attention.

Figure~\ref{fig:musk_refs} illustrates that the entrepreneurs exhibited a high frequency of mentions and hashtags twice. The first instance occurred in late April to early May, shortly after Mr. Musk's offer to purchase Twitter. The second instance occurred from late October to early November, following the finalization of the acquisition. Based on Figure~\ref{fig:musk_refs}, the initial surge in activity was brief, lasting only one week, and featured both mentions and hashtags being used equally excessively. In contrast, the second surge lasted approximately two months. It was primarily driven by direct mentions, which could be because the news of the completed acquisition was more personally significant to the users we observed, leading to more direct and personal responses.

It should be noted that the legal dispute in July 2022, in which Elon Musk declared his intention to terminate the acquisition, did not elicit a significant response in terms of the hashtag and mention frequencies. Specifically, there was no observable change in the frequency of mentions, and the increase in hashtag usage was minimal.

\subsection{Classifying the Users\label{sec:classification}}

Our goal is to identify subgroups of users that exerted different reactions to the acquisition process. We pursued the trajectory clustering approach proposed in~\cite{Andrienko09} to achieve this. Although the aggregate tweeting activity over the observation period was remarkably stable, both on average and by country (excluding the dip during the holiday season and the gradual decline in the first half of 2022), we believe that there might be hidden patterns in the tweeting behavior of subgroups of users that are relevant to our investigation. Therefore, by applying trajectory clustering, we aim to identify groups of users with similar tweeting patterns and analyze their reactions.

We computed a tweeting trajectory for each observed user by quantifying their tweeting frequencies at, a week and two weeks prior to, a week and two weeks after, and between the notable events (depicted as the ``dashed lines'')\footnote{Specifically, we used the following dates: 2022-03-31, 2022-04-07, 2022-04-14, 2022-04-21, 2022-04-28, 2022-06-25, '022-07-01, 2022-07-08, 2022-07-15, 2022-07-22, 2022-10-13, 2022-10-20, 2022-10-27, 2022-11-03, 2022-11-10, and 2023-01-29.}. Thus, each trajectory was represented as a 15-dimensional vector of real numbers. The averaging of the intervals between the acquisition events aimed to eliminate any sensitive response to unrelated episodes, such as national elections.

\begin{figure}[tb!]
\centering
\includegraphics[width=\textwidth]{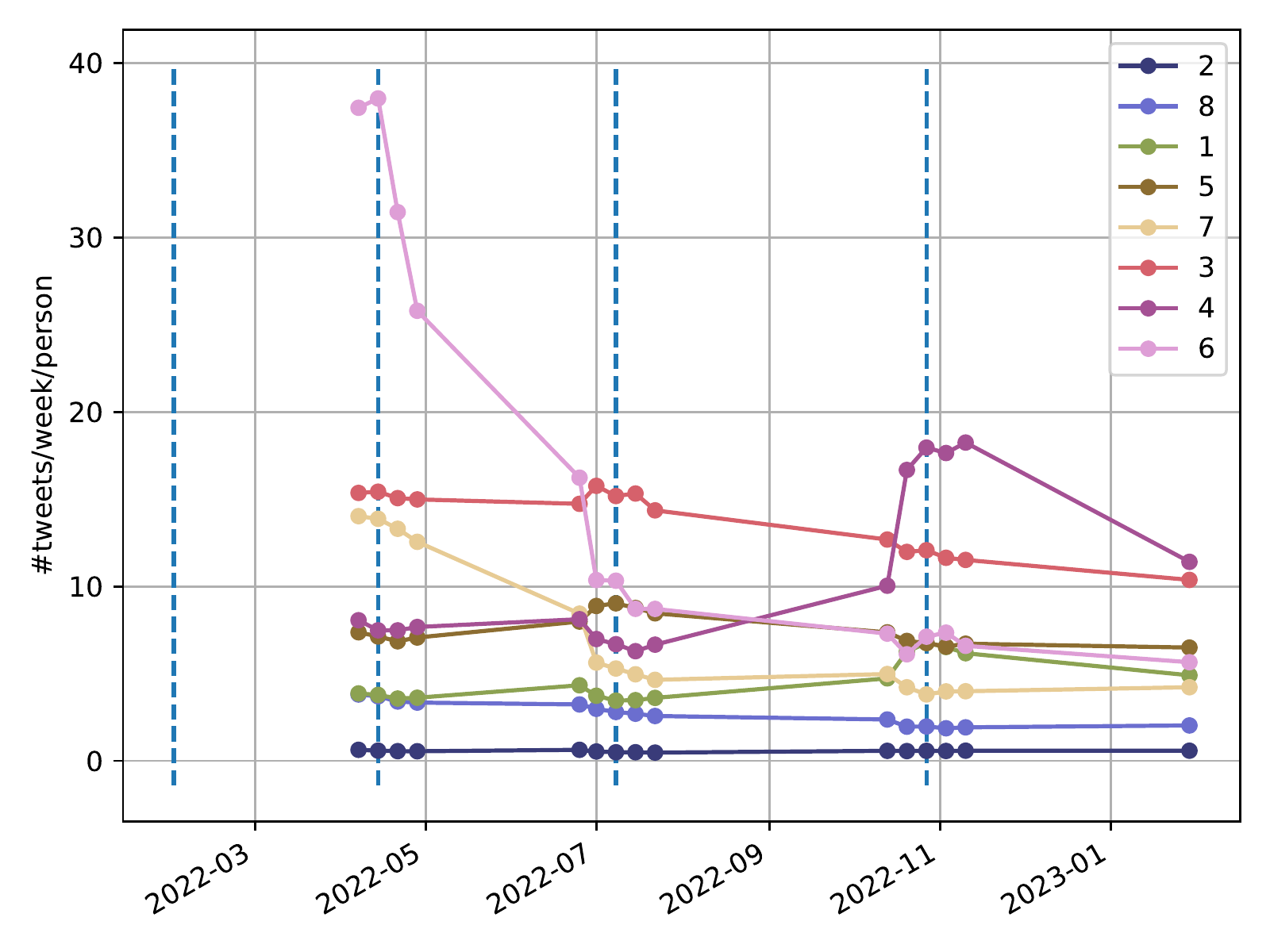}
\caption{Tweeting frequency per group over time. The groups are arranged in descending order of size. The point values represent the mean tweeting frequencies during the previous observation period.}
\label{fig:tag_trends}
\end{figure}

We further applied the k-means clustering algorithm~\cite{Jin2010} to divide the vectors into eight distinct and non-overlapping groups---clusters. Each vector in a cluster is closer (based on the Euclidean distance) to the vectors in the same cluster than those in any other cluster. The centroid, the arithmetic mean of all vectors in a cluster, is representative of that cluster. We call all users whose tweeting frequency trajectories belong to the same cluster, a group. Figure ~\ref{fig:tag_trends} shows the centroids of the eight groups obtained through this process, arranged in descending order of group size.

We selected the number of groups to be eight to ensure that the centroids of each group were not too similar and to separate users with abnormal behavior into distinct groups. Selecting any other number of groups would have violated one of these requirements.

Five of the eight groups consist of users with a stable tweeting history who were not significantly affected by the company acquisition events. The centroids of these groups differ only in terms of activity levels but are otherwise close to straight horizontal lines in ascending order: 2, 8, 1, 5, and 3. Interestingly, there is an inverse correlation between group size and activity level, suggesting that most Twitter users are passive observers or re-tweeters.

Groups 7 and 6, where the latter is the smallest, consisting only of 0.6\% of the dataset, exhibited reduced activity between the announcement of the intention to acquire Twitter and the attempted withdrawal from the deal. These groups are primarily responsible for the linear decline observed on the left side of Figure~\ref{fig:dynamics}.

Lastly, the users in group number 4 demonstrate a significant increase in activity in anticipation of and immediately after the closure of the acquisition deal, despite initially being aligned with (and potentially members of) group number 5. This group, although the second smallest (1.9\%), is the most vocal responder to the corporate leadership change.

\subsection{Extracting Hashtag-Based Conversation Topics}

Grouping the users into groups revealed a structural response of the entrepreneurs on Twitter to the takeover. On the contrary, topic analysis via hashtags addresses their semantic response, if any.

We utilized complex network analysis on a network of hashtags to identify the topics being discussed~\cite{zinoviev2018}. The network consisted of nodes representing the 3,000 most frequently used hashtags (\#ai, \#leadership, \#business, \#twitch, \#bitcoin, \#marketing, \#crypto, \#motivation, \#realestate, \#technology, etc.) 

An edge was created between two nodes if the corresponding hashtags were used in the same tweet at least once, and the weight of the edge was proportional to the number of simultaneous occurrences, normalized to the unit range. The resulting network had 2,428 distinct hashtags (including \#elonmusk) and 92,304 edges of various weights, and we removed edges between hashtags that co-occurred only once or twice. The ten strongest edges in the network were (\#ai, \#machinelearning), (\#realestate, \#realtor), (\#inspiration, \#motivation), (\#fashion, \#style), (\#digitalmarketing, \#marketing), (\#twitch, \#twitchstreamer), (\#ai, \#technology), (\#business, \#marketing), (\#bitcoin, \#crypto), and (\#tech, \#technology).

\begin{figure}[tb!]
\centering
\includegraphics[width=\textwidth]{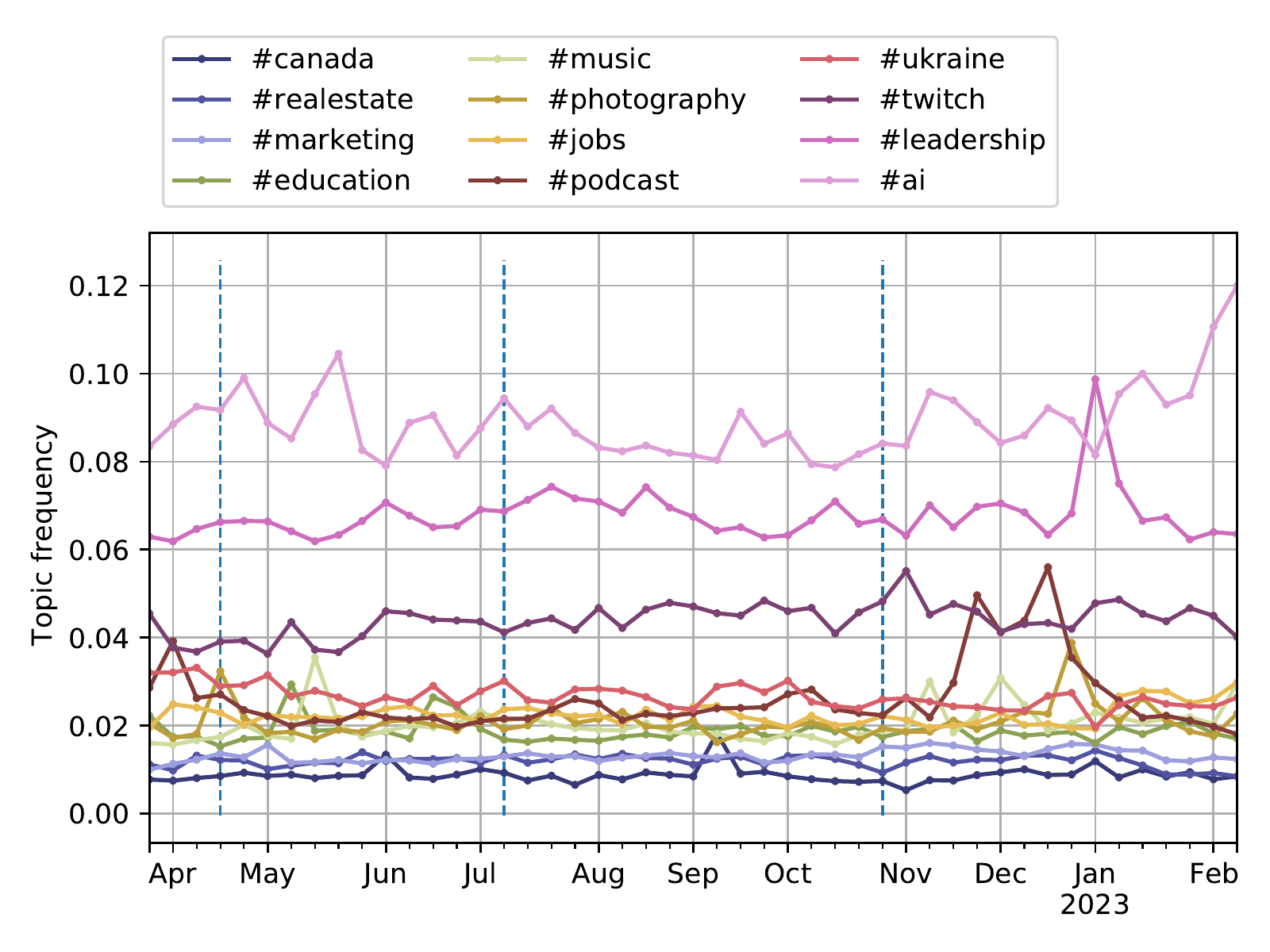}
\caption{Hashtag-based conversation topic frequencies over time.}
\label{fig:topic_trends}
\end{figure}

We employed the Louvain community detection method~\cite{blondel08} to identify groups of tightly connected network nodes representing hashtag-based conversation topics. This method produced several dozen topics, from which we selected the twelve most productive topics for further analysis, as shown in Figure~\ref{fig:topic_trends}. Each topic in the figure is labeled with its most frequently used hashtag, which may not fully reflect its contents. Therefore, we provided additional cues in Table~\ref{table:topics} by listing more hashtags associated with each topic.

\begin{table}[ht!]\centering
  \caption{\label{table:topics}The first six most frequent hashtags in the twelve most productive conversation topics. Both topics and hashtags within topics are listed in decreasing order of frequency.
  }\vskip0.5\baselineskip
  \begin{tabular}{ll}
    \hline
      {\bf \#ai} & \#bitcoin; \#crypto; \#technology; \#cybersecurity; \#auspol  \\
    {\bf \#leadership} & \#business; \#motivation; \#love; \#covid19; \#mentalhealth \\
     {\bf \#twitch} & \#tiktok; \#gaming; \#twitchstreamer; \#gamedev; \#youtube   \\
     {\bf \#ukraine} &  \#investing; \#southafrica; \#gold; \#energy; \#inflation\\
     {\bf \#podcast} &  \#writingcommunity; \#fifaworldcup; \#peace; \#books; \#amwriting\\
    {\bf \#jobs} & \#job; \#career; \#hiring; \#recruitment; \#hr\\
     {\bf \#photography} & \#travel; \#christmas; \#nature; \#runningwithtumisole; \#ireland \\
     {\bf \#music} & \#shopmycloset; \#poshmark; \#fashion; \#eurovision; \#loveisland \\
     {\bf \#education} & \#edtech; \#edchat; \#cdnpoli; \#inclusion; \#diversity   \\
   {\bf \#marketing} & \#socialmedia; \#digitalmarketing; \#twitter; \#contentmarketing; \ldots{} \#elonmusk    \\
    {\bf \#realestate} & \#realtor; \#homesforsale; \#property; \#toronto; \#florida\\
    {\bf \#canada} & \#australia; \#women; \#london; \#nyc; \#newyork   \\\hline
  \end{tabular}
\end{table}

Most topics depicted in Figure~\ref{fig:topic_trends} exhibit notably
stable dynamics except for three topics. Specifically, the surge in
the \#podcast and \#ai topics can be readily attributed to the FIFA
World Cup 2022 and the cryptocurrency exchange FTX implosion. Even
though the underlying reason for the growth of the \#leadership topic
remains uncertain, it cannot be linked to any Twitter acquisition
event. It implies that there was no observable conversation topic on
the entrepreneurs' Twitter accounts that reflected or was affected by
the takeover.

\subsection{Estimating Political Leaning\label{sec:leaning}}

The research question remaining to be investigated concerns the
potential influence of political leaning on entrepreneurs' reactions
to the change of ownership. A subnetwork of the entrepreneurs as
Twitter followers was extracted to examine this. It is assumed that if
a user follows another user, they share some views with the followed
user~\cite{burgess2020} and may lean politically in the corresponding
direction.

However, constructing a one-dimensional political spectrum scale
compatible with the liberal-conservative scale in the USA is a
challenging task due to the international nature of the
dataset. Creating such a scale necessitates a profound and
multifaceted understanding of local politics, which the authors
lack. Therefore, they have opted to use the existing North American
liberal-conservative scale as a universal yardstick, acknowledging the
potential for accusations of Americentrism while fully recognizing
that the scale may not fully capture local nuances.

\begin{figure}[b!]
\centering
\includegraphics[width=\textwidth]{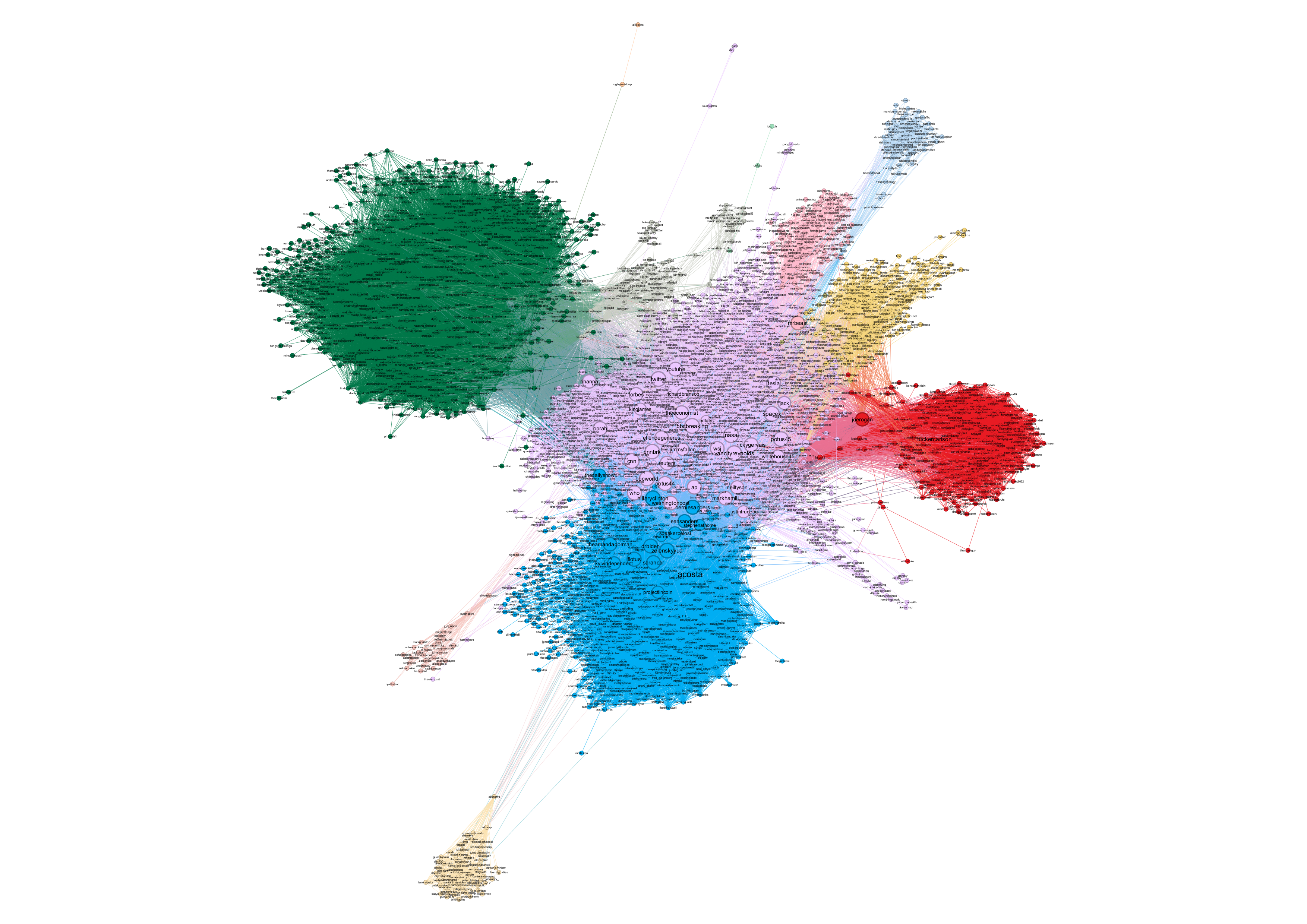}
\caption{The network of Twitter accounts based on co-following, with major network communities detected: ``liberal'' (blue), ``neutral'' (magenta), ``conservative'' (red), ``techno'' (yellow and pink), and ``South African'' (green). }
\label{fig:leadermap}
\end{figure}

For each user in the dataset, we collected a list of up to 1,000 users they follow. The number was limited by the recent changes in Twitter's API policies that severely restrict the followee lists' download speed.  A bipartite, directed, unweighted network was constructed with entrepreneurs as one set of nodes and their followees as the other set (referred to as ``opinion leaders,'' for lack of a better term). If multiple entrepreneurs followed the same two opinion leaders, a new edge was created connecting the leaders, thereby building a new projected network of co-followed~\cite{neal2014} (Figure~\ref{fig:leadermap}).

Some opinion leaders in the projected network are co-followed by liberal and conservative users because of their prominence. For example, a strong co-following edge was discovered between @joebiden and @realdonaldtrump, with the former signaling liberal-leaning and the latter indicating conservative-leaning. Since connected leaders may appear on the same followees list, they cancel out each other's contribution by sending controversial signals. The closeness centrality of the leaders' nodes can be used to estimate the propensity for true political signaling~\cite{freeman1979}. If a leader is close, in terms of the number of connecting edges, to the other leaders, they are at the center of the political cloud and do not cause much leaning in either direction. Conversely, a node with lower closeness centrality is away from the center and is politically biased. The leaders with the highest closeness centrality\footnote{Such that for the user $i$, the closeness centrality $x_i>\mu\left(x\right)+3\sigma\left(x\right)$.}, namely, @elonmusk, @joebiden, @barackobama, @potus, @kamalaharris, @realdonaldtrump, @aoc, @vp, @billgates, @nytimes, @michelleobama, @trevornoah, @whitehouse, @cnn, and @nasa, and their incident network edges were eliminated. The resulting network is sparser, has higher modularity, and has a crisper community structure~\cite{newman2006}.

Upon applying the Louvain community detection algorithm~\cite{blondel08}, we identified 14 groups of 2,491 opinion leaders, including highly politically polarized communities~\cite{conover2011}. The findings are summarized in Table~\ref{table:leaders}.

\begin{table}[ht!]\centering
  \caption{\label{table:leaders}The top opinion leaders in each cluster. Both clusters and leaders within clusters are listed in the decreasing order of the number of followers.
  }\vskip0.5\baselineskip
  \begin{tabular}{rrll}
  \hline
  \#&Size&Summary&Top members\\\hline
  1&    555,672 &Neutral& @spacex, @bbcbreaking, @gretathunberg, @vancityreynolds, \\
  &&&@twitter, @cnnbrk, @hillaryclinton, @bbcworld, @tesla\\
  2 &    286,231 &ZA& @cyrilramaphosa, @iamsteveharvey, @julius\_s\_malema,\\
  &&& @news24, @enca, @realblackcoffee, @presidencyza\\
  3  &   244,022 &Liberal& {\bf @zelenskyyua, @flotus, @drbiden, @sarahcpr,} \\
  &&&{\bf @speakerpelosi, @theamandagorman, @projectlincoln}\\
  4 &    168,054 & Entertainm't&@rihanna, @kingjames, @drake, @kevinhart4real, @espn,\\
  &&& @sportscenter, @kimkardashian, @snoopdogg, @iamcardib\\
  5  &   168,054 & Conservative&{\bf @potus45, @joerogan, @jordanbpeterson, @flotus45,}\\
  &&&{\bf  @foxnews, @benshapiro, @cobratate, @donaldjtrumpjr}\\
  
  6 &     122,086 &Tech/Crypto& @jack, @garyvee, @coinbase, @vitalikbuterin, @chamath,\\
  &&& @cz\_binance, @lexfridman, @naval, @binance, @sahilbloom\\
  7 &    79,380 & Social media& @mrbeast, @twitch, @playstation, @xbox, @pulte, @discord,\\
  &&&   @steam, @nintendoamerica, @elgato, @rockstargames\\
  8  &    48,193 & Soccer&@fabrizioromano, @cristiano, @premierleague, @f1,\\
  &&&@lewishamilton, @championsleague, @manutd\\
  9   &   28,617 & AU&@albomp, @abcnews, @danielandrewsmp, @senatorwong, \\
  &&&@latingle, @normanswan, @mrkrudd, @turnbullmalcolm\\
  10&     22,332 & IE& @donie, @rtenews, @leovaradkar, @simonharristd, \\
  &&&@irishtimes, @drtonyholohan, @gavreilly, @dublinairport\\
  11 &     15,196 & Business&@siriouslysusan, @cynthialive, @benlandis, \\
  &&&@murraynewlands, @shwood, @johnrampton, @digitaltrends\\
  
  12&      2,344 & Tennis&@rogerfederer, @djokernole, @wimbledon, @rafaelnadal\\
  13&      2,320 & Education&@edutopia, @googleforedu, @usedgov, @mindshiftkqed,\\
  &&&       @educationweek\\
  14 &      1,464 & Military& @usnavy, @usairforce, @usmc, @usarmy\\\hline
\end{tabular}

\end{table}

Of particular interest are clusters \#1 (the biggest and largely politically neutral, despite the presence of notable political figures),  \#3 (the ``liberal'' cluster), and \#5 (the ``conservative'' cluster). As a result, we can calculate the number of opinion leaders on lists \#3 and \#5 that each entrepreneur in the dataset follows and interpret the difference between the two as an indicator of their political leaning.

\begin{figure}[tb!]
\centering
\includegraphics[width=0.8\textwidth]{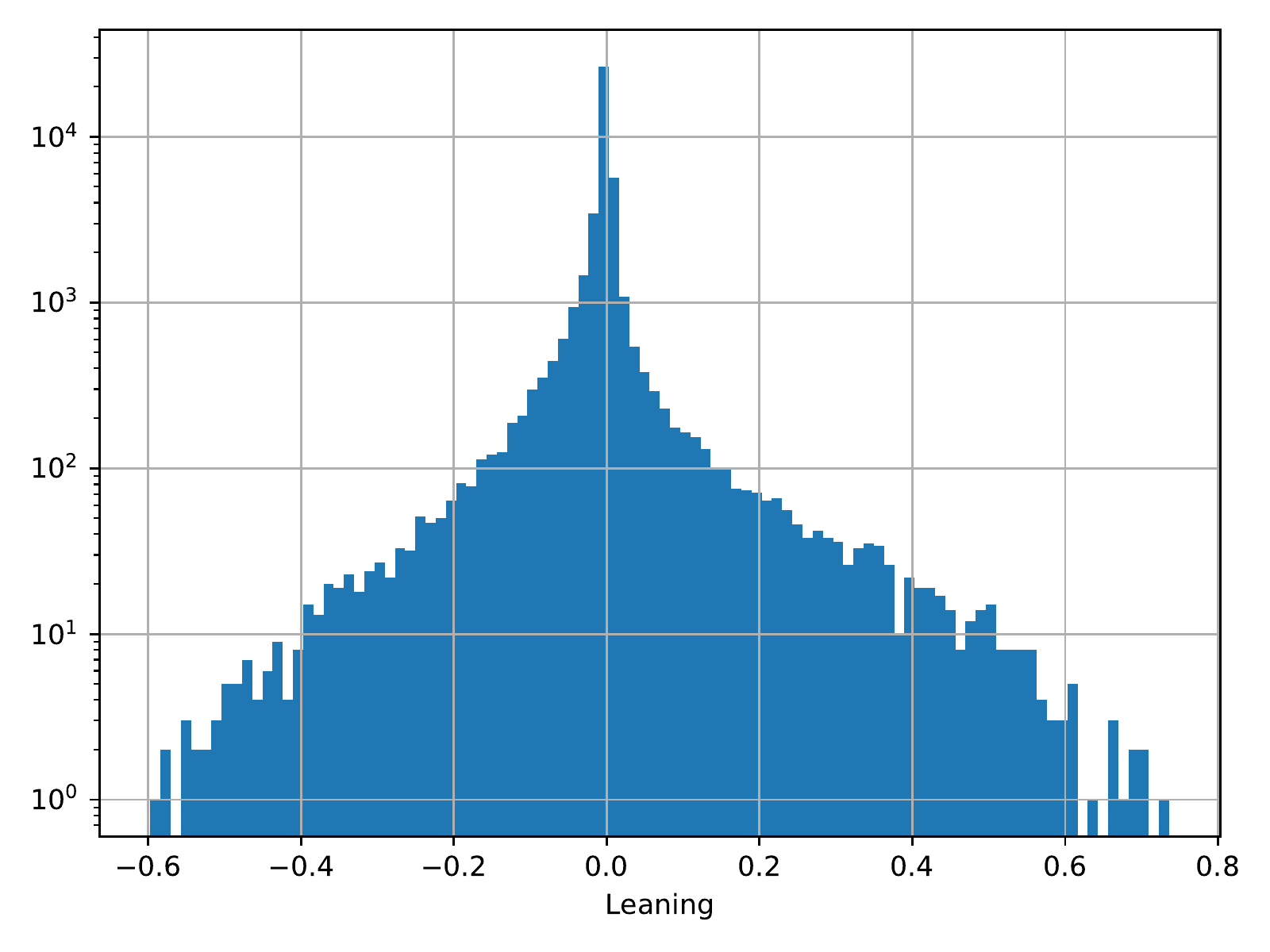}
\caption{Estimated political leaning of $\sim$45,500 users (``liberals'' on the left, ``conservatives'' on the right).}
\label{fig:leaning}
\end{figure}

Figure~\ref{fig:leaning} displays a histogram of the political leaning for all included entrepreneurs, revealing a symmetrical distribution with the mean and median leaning close to 0. The shape of the distribution suggests that, within the context of the proposed political leaning estimation framework, Twitter as a whole is politically neutral, at least within the sample of entrepreneurs.

\section{Discussion\label{sec:regression}}

We employed a logistic regression model to identify the factors that explain the entrepreneurs' association with the groups. Due to missing gender descriptions, only 32,654 users were included in the regression analysis. The results of this analysis are presented in Tables~\ref{table:regression} and~\ref{table:compare}.

\begin{table}[tb!]\centering
  \caption{\label{table:regression}Group sizes and logistic regression
    coefficients}\vskip0.5\baselineskip \small{\begin{tabular}{lrrrrrrrr}
  \hline
Groups:&2&8&1&5&7&3&4&6\\\hline
(Size, \%) & 57.4 & 18.6 & 7.7 & 7.5 & 3.3 & 3.0 & 1.9 & 0.6\\\hline
Male?    & -0.02 & -0.01 & 0.07 & -0.06 & 0.01 & 0.12$^{+}$ & 0.10 & 0.31$^{+}$\\
Leaning & 0.27 & 0.00 & -0.45 & 0.07 & -0.22 & -0.21 & 0.76 & -0.37\\
From CA? & -0.11$^{*}$ & 0.03 & 0.03 & 0.06 & 0.23$^{+}$ & 0.14 & 0.21 & -0.26\\
From IE? & 0.09$^{+}$ & 0.00 & 0.01 & -0.18$^{+}$ & -0.30$^{+}$ & 0.10 & -0.15 & -0.46\\
From NZ? & -0.10 & 0.14 & -0.09 & -0.26 & 0.20 & 0.37$^{+}$ & 0.38 & -0.48\\
From UK? & -0.08 & 0.10$^{+}$ & 0.07 & -0.01 & -0.35$^{*}$ & 0.26$^{*}$ & 0.20 & -1.16$^{**}$\\
From US? & -0.16$^{***}$ & 0.03 & -0.02 & 0.19$^{*}$ & 0.25$^{*}$ & 0.13 & 0.26$^{+}$ & -0.33\\
From ZA? & 0.25$^{***}$ & -0.21$^{***}$ & -0.26$^{**}$ & -0.16 & -0.08 & -0.19 & 0.14 & 0.09\\
Account age & 0.09$^{***}$ & -0.03$^{***}$ & -0.04$^{***}$ & -0.04$^{***}$ & -0.10$^{***}$ & -0.11$^{***}$ & -0.1$^{***}$ & -0.2$^{***}$\\
Descr. length & -0.57$^{***}$ & 0.31$^{***}$ & 0.40$^{***}$ & 0.33$^{***}$ & 0.40$^{**}$ & 0.63$^{***}$ & 0.62$^{**}$ & -0.20\\
Has URL? & -0.47$^{***}$ & 0.18$^{***}$ & 0.30$^{***}$ & 0.41$^{***}$ & 0.37$^{***}$ & 0.51$^{***}$ & 0.30$^{***}$ & 0.29$^{+}$\\
log(\#followers) & -0.26$^{***}$ & 0.20$^{***}$ & 0.2$^{***}$ & -0.09 & 0.08 & 0.23$^{**}$ & 0.07 & 0.39$^{**}$\\
log(\#following) & 0.37$^{***}$ & -0.09$^{*}$ & -0.16$^{**}$ & -0.31$^{***}$ & -0.08 & -0.41$^{***}$ & -0.27$^{**}$ & -0.50$^{**}$\\
log(\#listed) & -0.03 & -0.05 & 0.05 & 0.06 & -0.19$^{**}$ & 0.14$^{+}$ & 0.06 & 0.04\\
log(\#tweets) & -1.25$^{***}$ & 0.34$^{***}$ & 0.55$^{***}$ & 1.23$^{***}$ & 1.18$^{***}$ & 1.48$^{***}$ & 1.28$^{***}$ & 1.57$^{***}$\\
Verified? & -0.26$^{**}$ & -0.04 & -0.01 & 0.15 & 0.24 & -0.06 & 0.36$^{+}$ & -0.73\\
\hline
\end{tabular}
}
\end{table}

The first insight from Table~\ref{table:regression} is that except
Group 2, all other groups have relatively newer Twitter accounts,
which shows that they have not been on Twitter that long. We remind
the readers that Group 2 is the biggest of all the groups and
constitutes about 57.5\% of all the entrepreneurs in our
study. With that in mind, it will be interesting to not only
examine how some of the smaller groups are similar to and different
from this majority group in terms of their characteristics but also
observe what makes these smaller groups similar to and different from
each other.

Interestingly, we find that Group 1 and Group 8 have similar traits on
several dimensions:

\begin{itemize}
  \item Entrepreneurs who belong to both of these groups are not from
    South Africa;
  \item They provide fairly long bio descriptions, including their
    personal website URL, that shows that they want to be known and
    want people to land on their webpage for more information about
    themselves and what they do;
  \item They both have strong followers but entrepreneurs belonging to
    either of these groups do not necessarily follow a high number of
    accounts themselves (as captured by the ``Following'' count from
    their respective Twitter profiles); and,
  \item They both have a relatively high frequency of tweets.
\end{itemize}

Based on the above set of common characteristics, we believe that
entrepreneurs belonging to Groups 1 and 8 may best be considered
opinion providers who use Twitter for the interest of self/business
promotion. In other words, these enterpreneurs' Twitter presence seems
to be largely business-driven.

Entrepreneurs in Group 3 have many traits similar to those in Groups 1
and 8, except for where they are from. They, too, have relatively
newer Twitter accounts, long bio descriptions including personal URLs,
strong followers but not necessarily following many accounts
themselves, and a high frequency of tweets. What separates Group 3
from Groups 1 and 8 is that they are primarily located in the UK and New
Zealand. Also, this group has significantly more male
entrepreneurs.

Next, looking at Group 6, we see that this group also has a strong
male presence with a high number of tweets with a short description
and without URLs, and consists of a disproportionately lower number of
people from the UK. Its members have relatively newer account on
Twitter and strong followers but do not necessarily follow a high
number of accounts themselves. These features indicate that those
entrepreneurs provide opinions to their followers but not necessarily
want to be known or interested in people knowing who they are or what
they do. This trait of entrepreneurs in Group 6 is similar to those in
Groups 1, 3, and 8, who are more into being known/promoting themselves
and their businesses. What mainly separates these groups are mostly
where they are located. We call them the {\em self-promoters}.

Continuing to examine our regression results further, we see that
Groups 4, 5, and 7 stand out a different spectrum. While being similar
to each other and some of the earlier groups in terms of having
relatively newer Twitter with long bio descriptions, including their
personal URL and high tweet frequency, enterpreneurs who are part of
this group neither have many followers and follow many other Twitter
accounts themselves nor necessarily have many followers (i.e.,
the estimated coefficients were statistically insignificant). We
decided to label them as the {\em wannabe
  self-promoters}---entrepreneurs who are on their path to becoming
self-promoters one day, but do not have the critical mass of
followers qualifying them as self-promoters. Also, we notice
that entrepreneurs from Groups 4, 5, and 7 mostly seem to
originate from the US.

Finally, let us examine Group 2---our most populous group of
entrepreneurs. Besides accommodating an overwhelming majority of our
entrepreneurs (57.5\%) in our study, this group has a set of starkly
distinct features that collectively help entrepreneurs in this group
stand out from all the other groups.

For example, the entrepreneurs belonging to Group 2 have relatively
older Twitter accounts, a relatively shorter bio descriptions with no
URL (indicating that these entrepreneurs are possibly not keen on
self/business promotional activities), and do not have a high number
of followers. However, they follow a relatively high number of
accounts and have the distinctive trait of not tweeting much! In other
words, entrepreneurs in our most populous group are on Twitter mostly
to silently read and observe. This group also constists of a
significantly large proportion of entrepreneurs who do not have a
verified account, i.e., they do not have the blue check associated
with their Twitter accounts\footnote{The blue verified badge on
  Twitter lets people know that an account of public interest is
  authentic.}. That means these people are not interested in being
publicly known by promoting the authenticity of their Twitter
accounts. We call them the {\em silent observers}.

Overall, entrepreneurs in Groups 1, 3, 6, and 8, whom we broadly label
as the {\em self-promoters}, are regularly on Twitter to be
known and use the platform for self/business interests. Twitter is
mostly a marketing tool for them to connect with their target
audience and find new audiences.

Collectively, they comprise about 29.9\% of the sample we studied. We
label the entrepreneurs belonging to Groups 4, 5, and 7 as the {\em
  wannabes} and believe they have the same intention as the
self-promoters but are not quite there yet. They make up about 13.3\%
of the sample that we studied. Interestingly, entrepreneurs belonging
to only Group 4 seem to own a relatively higher proportion of
blue-check verified Twitter accounts. About 43.2\% of the
entrepreneurs we studied use the Twitter platform for self-promotion
or want to be linked/known for who they are and what they do, while a
larger number (57.5\%) of entrepreneurs use Twitter just to read and
observe silently.

\begin{table}[tb!]\centering
  \caption{\label{table:compare}Average statistics by groups with
    similar tweeting trajectories}\vskip0.5\baselineskip
  \small{\begin{tabular}{lrrrl}
  \hline
Groups:&A=1+2+3+5+8&B=6+7&C=4&Student's t-test\\\hline
Male?&0.66 & 0.65&0.66&ns\\
Leaning & -0.001 & -0.002 & 0.001 & ns\\
Account age&9.04&9.27&9.51&A$\ne$B$^{***}$, A$\ne$C$^{***}$\\
Descr. length&0.70&0.74&0.77&A$\ne$B$^{***}$, A$\ne$C$^{***}$, B$\ne$C$^{***}$\\
Has URL?&0.55&0.64&0.75&A$\ne$B$^{***}$, A$\ne$C$^{***}$, B$\ne$C$^{***}$\\
log(\#followers)&2.61&2.83&3.09&A$\ne$B$^{***}$, A$\ne$C$^{***}$, B$\ne$C$^{***}$\\
log(\#following)&2.71&2.80&2.86&A$\ne$B$^{***}$, A$\ne$C$^{***}$, B$\ne$C$^{**}$\\
log(\#listed)&0.77&0.95&1.21&A$\ne$B$^{***}$, A$\ne$C$^{***}$, B$\ne$C$^{***}$\\
log(\#tweets)&3.28&3.54&3.86&A$\ne$B$^{***}$, A$\ne$C$^{***}$, B$\ne$C$^{***}$\\
Verified?&0.02&0.03&0.05&A$\ne$B$^{***}$, A$\ne$C$^{***}$, B$\ne$C$^{**}$\\
\hline
\end{tabular}
}
\end{table}

Next, in Table~\ref{table:compare}, we report the summary statistics of
our entrepreneurs by further grouping them (i.e., forming
groups of groups) based on the broad similarity of their tweeting
trajectories (see Figure~\ref{fig:tag_trends}).

Entrepreneurs in Group 4, the {\em wannabe self-promoters}, easily
stand out and seem to have the most distinct change in their tweet
trajectoryin response to Elon Musk's official acquisition of Twitter
(November 2022). We find that, just before the acquisition deal was closed,
their tweet frequency increased significantly, continued increasing
for a little but post-acquisition, had a flat trajectory for a very
short while after closing, and then declined sharply thereafter. In
sharp contrast to the overall muted sentiments from entrepreneurs
belonging to most other groups in response to the official
acquisition event, this group showed the most visible reaction to
Musk's acquisition of Twitter.

Knowing they are wannabes and want to self-promote and tweet with high
frequency, their strong reaction (positive or negative) makes
sense. Groups 6 and 7 (also wannabes), both with a desire to
have strong followers, had high tweet frequency when Musk announced
his intention to buy Twitter (May 2022). However, their tweeting frequency
declined sharply after that and went almost flat when Musk announced
his intention to withdraw from the deal (July 2022). Both had a
significantly noticeable instantaneous reaction to the final official
acquisition, which eventually became flat. For all the other remaining
groups (1, 2, 3, 5 and 8), the tweeting trajectory is very similar
over our entire observation period. Though these groups vary
significantly from each other in terms of their frequencies of
tweeting, their overall reaction to Musk buying Twitter is quite the
same---and can be considered very muted at best!

In our study, we observe and analyze the tweeting behavior of more
than 32,000 entrepreneurs who seem to sort them into eight distinct
groups based on their motives, intentions, and usage of the Twitter
platform. Surprisingly, we do not find any significant difference
between the groups in terms of their political affiliation (left-wing
or right-wing). When we look at their behavioral tendencies, we see
that some are closer to one another: Groups 1, 3, and 8 as the {\em
  self-promoters}; Groups 4, 5, 6, and 7 as the {\em wannabes}; and
Group 2 as {\em silent observers}. Irrespective of their diverse
intentions and motives to be on Twitter and their varying frequency of
platform usage, they all have similar reactions to Elon Musk’s
acquisition of Twitter, i.e., no reaction---which we thus call a
``Non-Musk Effect.'' We believe that the instantaneous reaction that
we observe in only a few of the groups (Groups 4, 6, and 7), also
conform with the behavior pattern of the other groups, as all of their
tweeting trajectories went flat very soon. In the Twitterverse,
entrepreneurs quickly returned to their world or never left it in most
cases, even as Musk's antics vis-à-vis the Twitter acquisition deal
had caught the popular press and technology world in a frenzy!

\section{Conclusion}
After months of speculation and drama on buying or not buying Twitter, the widely popular social media app was finally sold to Tesla CEO Elon Musk in a deal of \$44 billion in November 2022. While the purchase gave Musk control of another giant company, many people have had mixed feelings about what that means for the future of Twitter. Media reaction was extensive during the negotiation deals, during the purchase, and even after the purchase. Survey companies' extensive surveys also showed a split in thoughts on what this deal may mean. One survey of over 2,000 Americans by Preply.com found that 2 in 3 people did not want Musk to take control of Twitter~\cite{zajechowski2022}. In another survey by platform OnePulse, 55.4\% of 1,000 Americans believe Musk has his interests in mind about buying Twitter~\cite{elonmusktwittertakeoveroffer2023}.

Since taking over, the turmoil within Twitter has only got more escalated. Musk has cut Twitter's staff by more than two-thirds and engaged in an aggressive campaign of cost-cutting targeting Twitter's workplace and infrastructure, reorienting the company around what he regards as his star team of engineers. Before taking over, Musk had heaped criticism on his predecessors' approach to content moderation, casting them as a pro-censorship regime stifling free speech. Meanwhile, his team has rolled out a system of different-colored verification check marks to distinguish between government, business, and other verified accounts.

With all that turmoil since October 2022, did the Twitter users care
that much about Musk's take over as the mass media portrayed? This
research seeks to fill this gap by examining the ``Musk Effect'' on
the Twitter activity of self-declared entrepreneurs. It quantifies the
``Musk Effect'' on the tweeting activity of entrepreneurs from seven
English-speaking countries (US, Australia, New Zealand, UK, Canada,
South Africa, and Ireland) for 71 weeks. We downloaded and created a
dataset of 9.9 million tweets from more than 47,000 self-declared
entrepreneurs. We used trajectory cluster analysis to classify the
entrepreneurs who led to eight broad non-overlapping
groups. Hashtag-based analysis based on extracting popular
conversation topics from these entrepreneurs' tweets showed that there
was no observable conversation topic on the entrepreneurs' Twitter
accounts that reflected or was affected by Musk's Twitter
takeover. Most topics were trending during the acquisition, such as
AI, cryptocurrency exchange, FTX implosion, etc. We also showed that
political leaning does not significantly affect users' attitudes
toward the Twitter acquisition process.

Results further revealed that less than 4\% of entrepreneurs react to Musk's acquisition news, which is signaled by these accounts' tweet frequency. That means the media reaction to Twitter's acquisition by Musk was not necessarily aligned with main street's reaction. People, in our case, entrepreneurs, did not care that much about Musk's ownership of Twitter. Most entrepreneurs, around 57.5\%, use the platform to read and observe. 12.6\% of them are there to promote their business or for self-promotion. Only less than 1\% are on Twitter to influence without any personal linkage.

All in all, if there was an effect of Musk's acquisition of Twitter among entrepreneurs, it was a non-Musk effect!

\section{Acknowledgments}
The authors are grateful to Bari Bendell for the participation in the dataset definition.

\bibliographystyle{acm}
\bibliography{cs}

\end{document}